\documentclass[aps,prb,twocolumn]{revtex4}
\begin{document}

\title{Disordered Bose-Einstein Condensates in Quasi One-Dimensional\\
Magnetic Microtraps}
\author{Daw-Wei Wang, Mikhail D. Lukin, and Eugene Demler}
\date{\today}

\begin{abstract}
We analyze the effects of a random magnetic potential in a microfabricated
waveguide for ultra-cold atoms. We find that the shape and position
fluctuations of a current carrying wire induce a strong Gaussian correlated
random potential with a lengthscale set by the atom-wire separation. The
theory is used to explain quantitatively the observed fragmentation of the
Bose-Einstein condensates in atomic waveguides. Furthermore, we show that
nonlinear dynamics can be used to provide important insights into the nature
of a quantum phase transition from the superfluid to the insulating Bose
glass phase may be reached and detected under the realistic experimental
conditions.
\end{abstract}

\pacs{PACS numbers: 05.30.Jp,03.75.Lm,67.40.Db}
\maketitle

\draft

\address{Physics Department, Harvard University, Cambridge, MA 02138}

%%%%%%%%%%%%%%%%%%%%%%%%%%%%%%%%%%%%%%%%%%%%%%

%%%%%%%%%%%%%% Introduction: microtrap %%%%%%%%%%%%%%%

Several groups have recently reported realizations of quasi one-dimensional
(Q1D) Bose-Einstein condensates (BEC) in the magnetic waveguides for atoms
created by the microfabricated wires (magnetic traps on microchips) [%
\onlinecite{microchip_general,kraft02,aaron1,zimmermann1}]. A surprising
feature is the presence of fragmentation: a random modulation of the atomic
density in the axial direction [\onlinecite{kraft02,aaron1,zimmermann1}].
For smaller atom-wire separation the fragmentation is enhanced and its
characteristic lengthscale decreases. Determination of the origin and
properties of the observed density modulation is essential for understanding
atomic condensates on microchips and for developing atom-optical
microfabricated devices. Several observations suggest that the wire shape
and position fluctuations are the primary origin of the fragmentation [%
\onlinecite{kraft02,aaron1,zimmermann1}], however, a detailed theoretical
understanding of these phenomena is still lacking [\onlinecite{henkel}].

In this Letter we use a first-principle microscopic calculation to
demonstrate that a small meandering of the wire leads to a strong random
potential in a magnetic microtrap [\onlinecite{note0}]. We show that even
when the wire fluctuations have no intrinsic lengthscale, the disorder
magnetic potential in the waveguide is a Gaussian correlated random
potential with a correlation function that is strongly peaked at a finite
wavevector and vanishes for small and large wavevectors. The characteristic
length scale of this disorder correlation function is set by the atom-wire
distance, $d$. This theoretical model is used to quantitatively explain the
experimental results. We then study the properties of the Q1D atomic BEC in
the regime of strong fragmentation, which may be realized in microtraps by
bringing atoms closer to the chip surface and/or changing the atomic
density. Some of the most striking manifestations of this strong disorder
limit appear in the non-linear dynamics of the fragmented condensate,
including a chaos and a self-trapping of the excitations. We argue that the
quantum phase transition between the superfluid (SF) and the insulating Bose
glass (BG) phases can be achieved and detected under realistic experimental
conditions.

\begin{figure}[tbp]
\vbox to 4.5cm {\vss\hbox to 5.cm
 {\hss\
   {\includegraphics{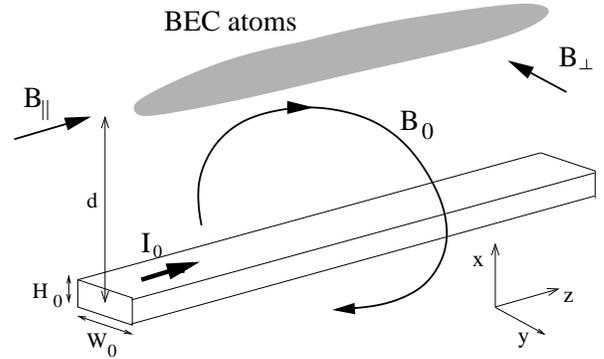}
   }
  \hss}
 }
\caption{Typical setup of a magnetic microtrap: A current $I_0$ flowing in a
microfabricated copper conductor and a perpendicular bias field, $B_\perp$,
form a magnetic waveguide for atoms. An offset field $B_\|$ applied parallel
to the wire reduces loss processes near the magnetic field minimum. }
\label{fig1}
\end{figure}

Neutral atoms with a magnetic dipole moment, $\mathbf{\mu}_a$, anti-aligned
with respect to the magnetic field, $\mathbf{B}$, experience a potential $V(%
\mathbf{r})=\mu_a |\mathbf{B}(\mathbf{r})|$, so they can be confined near a
field minimum [\onlinecite{microchip_general}]. A typical microtrap setup is
shown in Fig. \ref{fig1}. The waveguide is located at a distance $d=\frac{%
2I_0}{cB_\perp}$ from the wire center, where the azimuthal magnetic field
created by the wire cancels out the transverse bias field, $B_\perp$. The
longitudinal confinement can be provided by the additional magnetic field
gradients applied parallel to the wire.

%-------------
%----------------------

We now discuss a random magnetic potential along the center of the waveguide
caused by the shape and position fluctuations of the current carrying wire.
Let $f_{L/R}(z)$ be the deviations of the wire's left/right boundaries from
their ideal positions at $y=\mp W_{0}/2$. Changes in the wire height provide
a much smaller contribution to the fluctuating magnetic field, so we will
neglect them throughout this paper. In a steady state the deviation of the
current density, $\delta \mathbf{J}(y,z)$, from its average value, $J_{0}$,
satisfies the charge conservation condition, $\partial \delta J_{y}/\partial
y+\partial \delta J_{z}/\partial z=0$. If we simultaneously assume a
constant conductivity throughout the wire, the current is also vorticity
free, $\partial \delta J_{z}/\partial y-\partial \delta J_{y}/\partial z=0$.
Boundary conditions for $\delta \mathbf{J}(y,z)$ follow from the requirement
that no current flows out of the wire. Assuming small fluctuations of the
wire ($|f_{L/R}(z)|\ll W_{0}$ and $|\frac{\partial f_{L/R}}{\partial z}|\ll 1
$), we have $\delta J_{y}(\mp W_{0}/2,z)/{J_{0}}=\partial
f_{L/R}(z)/\partial z$. It is convenient to introduce an auxiliary scalar
potential $G_{J}(y,z)$, such that $\delta J_{y}=J_{0}\partial G_{J}/\partial
z$ and $\delta J_{z}=-J_{0}\partial G_{J}/\partial y$. The function $%
G_{J}(y,z)$ satisfies the Laplace equation in the interior of the wire and
has Dirichlet boundary conditions at $y=\mp W_{0}/2$. Using separation of
variables, we find 
\begin{eqnarray}
G_{J}(y,z) &=&\frac{2}{\pi }\int \frac{e^{ikz}dk}{\sinh (kW_{0})}\left[
\cosh (ky)\sinh (kW_{0}/2)f_{k}^{+}\right.   \nonumber \\
&&+\left. \sinh (ky)\cosh (kW_{0}/2)f_{k}^{-}\right] ,  \label{G_J}
\end{eqnarray}%
where $f_{k}^{\pm }\equiv \frac{1}{2}\int_{-\infty }^{\infty
}dze^{ikz}(f_{R}(z)\pm f_{L}(z))$ are the Fourier components of the wire
position and width fluctuations. The fluctuating part of the magnetic field
can be found from the Biot-Savarat law: 
\begin{eqnarray}
\delta \mathbf{B}(\mathbf{r}) &=&\frac{J_{0}}{c}\int_{-\infty }^{\infty
}dz^{\prime }\int_{-\frac{H_{0}}{2}}^{\frac{H_{0}}{2}}dx^{\prime }\int_{-%
\frac{W_{0}}{2}}^{\frac{W_{0}}{2}}dy^{\prime }G_{J}(y^{\prime },z^{\prime })
\nonumber \\
&&\times \left\{ -\frac{2(x-x^{\prime })^{2}-(y-y^{\prime
})^{2}-(z-z^{\prime })^{2}}{\left( (x-x^{\prime })^{2}+(y-y^{\prime
})^{2}+(z-z^{\prime })^{2}\right) ^{5/2}}\hat{x}\right.   \nonumber \\
&&+\frac{-3(x-x^{\prime })(y-y^{\prime })\hat{y}}{\left( (x-x^{\prime
})^{2}+(y-y^{\prime })^{2}+(z-z^{\prime })^{2}\right) ^{5/2}}  \nonumber \\
&&+\left. \frac{-3(x-x^{\prime })(z-z^{\prime })\hat{z}}{\left( (x-x^{\prime
})^{2}+(y-y^{\prime })^{2}+(z-z^{\prime })^{2}\right) ^{5/2}}\right\} ,
\label{delta_B_rec}
\end{eqnarray}%
To obtain Eq. (\ref{delta_B_rec}) we performed the integration by parts,
which led to a cancellation of the contributions from the $y\in (-\frac{W_{0}%
}{2}+f_{L}(z),-\frac{W_{0}}{2})$ and $y\in (\frac{W_{0}}{2},\frac{W_{0}}{2}%
+f_{R}(z))$ regions. The leading order random potential comes from the $z$%
-component of the magnetic field, $\delta U(\mathbf{r})=\pm \mu _{a}\delta
B_{z}(d,0,z)$, if the condensate width is much smaller than $d$ (the $\pm $
sign is the orientation of the offset field, $B_{\Vert }$). From Eqs. (\ref%
{G_J})-(\ref{delta_B_rec}) one can see that only the wire position
fluctuations, $f^{+}$, contribute to $\delta U(\mathbf{r})$ at this order.

%-----------
\begin{figure}[tbp]
\vbox to 7.5cm {\vss\hbox to 5.cm
 {\hss\
   {\includegraphics{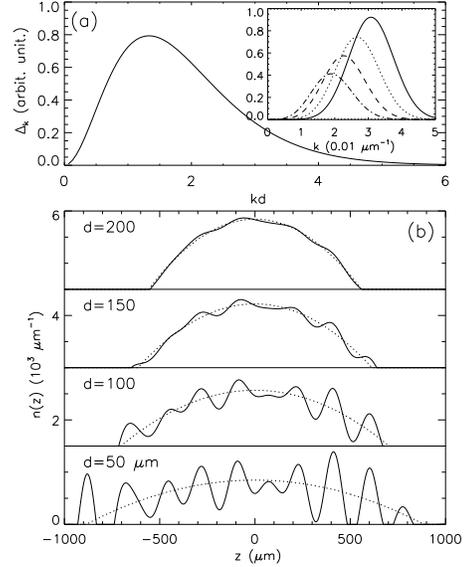}
   }
  \hss}
 }
\caption{ (a) Correlation function of the random magnetic potential in a
microtrap, $\Delta _{k}$, assuming white noise fluctuations of the wire
position (i.e. $F_{k}=$ constant). Insert: $\Delta _{k}$ when the wire
fluctuations have an intrinsic length scale: $F_{k}\propto (e^{-(k-k_{1})^{2}%
\protect\eta _{1}^{2}}+e^{-(k+k_{1})^{2}\protect\eta _{1}^{2}})$. We take $2%
\protect\pi /k_{1}=200\protect\mu $m, $\protect\eta _{1}=100\protect\mu $m,
and $\langle f^{+}(z)f^{+}(z)\rangle _{\mathrm{dis}}=(0.1$ $\protect\mu $m)$%
^{2}$. Solid to dash-dotted lines correspond to $d=50$, 100, 150, and 200 $%
\protect\mu $m respectively. (b) Condensate density profiles with parameters
chosen to be close to the values used in the experiments of Ref. [
\onlinecite{aaron1}]. Dotted lines are the results \textit{without} random
potential.}
\label{u_x}
\end{figure}
%----------------
To elucidate the nature of the resulting random potential we compute the
correlation function $\Delta _{k}=\int dz\,e^{ik(z-z^{\prime })}\langle
\delta U(z)\delta U(z^{\prime })\rangle _{\mathrm{dis}}$, where $\langle
\cdots \rangle _{\mathrm{dis}}$ is disorder ensemble average. Assuming $%
H_{0}\ll d$ (this condition is typically satisfied in experiments), we
obtain 
\[
\Delta _{k}=\left( \frac{2I_{0}\mu _{a}}{c}\right) ^{2}\frac{(kd)^{4}}{d^{4}}%
|K_{1}(kd)D(kW_{0},kd)|^{2}F_{k},
\]%
where $F_{k}=\int dz\,e^{-ik(z-z^{\prime })}\langle f^{+}(z)f^{+}(z^{\prime
})\rangle _{\mathrm{dis}}$, and 
\begin{eqnarray}
D(x,y) &=&\frac{2\sinh (x/2)}{x\sinh (x)K_{1}(y)}\sum_{n=0}^{\infty }\frac{%
(-1)^{n}}{n!(2y)^{n}}K_{n+1}(y)  \nonumber \\
&&\times \left[ \gamma (2n+1,x/2)-\gamma (2n+1,-x/2)\right] .
\end{eqnarray}
$K_{n}(x)$ is the Bessel function of the second kind, and $\gamma
(n,x)\equiv \int_{0}^{x}dx^{\prime }x^{\prime }{}^{n-1}e^{-x^{\prime }}$ is
the incomplete Gamma function. The strength of the disorder potential can be
defined as $u_{s}^{2}\equiv \langle \delta U(z)\delta U(z)\rangle _{\mathrm{%
dis}}=\int_{-\infty }^{\infty }\frac{dk}{2\pi }\Delta _{k}$. For $d\gg W_{0}$
we obtain 
\begin{eqnarray}
u_{s}\sim \frac{2I_{0}\mu _{a}}{cd}\cdot \left( \frac{F_{0}}{d^{3}}\right)
^{1/2}.
\label{u_s}
\end{eqnarray}
Fixing current $I_{0}$ we find $u_{s}\sim d^{-5/2}$, which is close to the
experimentally observed, $u_{s}\sim d^{-2.2}$ [\onlinecite{kraft02}].We can
also estimate the magnitude of the random potential: If $\delta f\sim 0.1$ $%
\mu $m is the scale of the wire position fluctuation and $\xi \sim 100$ $\mu 
$m is the appropriate correlation length, then $u_{s}\sim \mu _{a}B_{\perp }%
\frac{\delta f\,\xi ^{1/2}}{d^{3/2}}\sim $ kHz for $B_{\perp }\sim 1$ G and $%
d\sim 100$ $\mu $m. This value is comparable to a typical value of the BEC
chemical potential and explains the strong fragmentation effects observed in
magnetic microtraps [\onlinecite{kraft02,aaron1,zimmermann1,note0}].

In Fig. \ref{u_x}(a) we show $\Delta _{k}$ computed for the white noise
fluctuations of the wire center position \ It peaks at $k_{0}\approx 1.3/d$
and vanish in the $k=0$ and $k=\infty ,$ showing a characteristic
lengthscale, $d$. Long wavelength fluctuations are suppressed because a
uniform shift of the random potential does not lead to a parallel ($z$)
component of the magnetic field. In the insert of Fig. \ref{u_x}(a), we
demonstrate how the random magnetic potential is modified when the wire
fluctuations have an intrinsic lengthscale $1/k_{1}$ (see caption). When the
wire fluctuations are short ranged ($k_{1}d\gg 1$), the potential
fluctuations remain peaked close to $k_{0}\sim 1.3/d$, and the lengthscale
of the random potential is set by the atom-wire separation. When the wire
fluctuations are longer ranged ($k_{1}d<1$), $\Delta _{k}$ is then peaked at 
$k\sim k_{1}$, and the random potential tracks the wire fluctuations. A
linear relation between the condensate height and the lengthscale of the
fragmentation was observed in Ref. [\onlinecite{aaron1,zimmermann1}]. 
In Fig. \ref{u_x}(b) we show the calculated condensate density profiles
for $10^{6}$ sodium atoms in the Thomas-Fermi approximation using the same
wire fluctuations taken from the distribution of the insert of Fig. \ref{u_x}%
(a). The fragmentation appears for $d\leq 100$ $\mu $m (c.f. Ref. [%
\onlinecite{aaron1}]). We note that the profiles obtained by a white noise
potential (i.e. $\Delta _{k}=$const.) do not have a characteristic length
scale, and are very \textit{different} from the above results and
experiments.

%---------------------------
\begin{figure}[tbp]
\vbox to 5.5cm {\vss\hbox to 5.cm
 {\hss\
   {\includegraphics{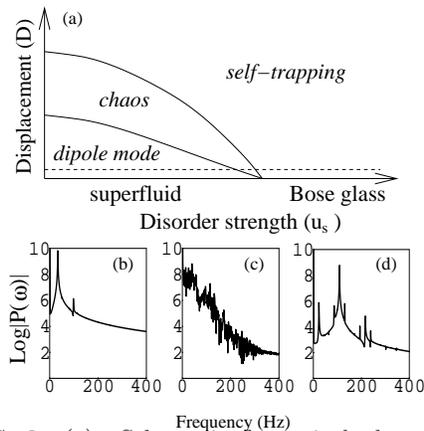}
   }
  \hss}
 }
\caption{(a): Schematic dynamical phase diagram of fragmented quasi
one-dimensional BEC in relation to the shaking experiments. The dashed line
indicates the minimum displacement to generate an observable density
fluctuations (i.e. $\protect\zeta>\protect\zeta_{\mathrm{min}}$). Three
types of the system response can be identified from the power spectra of the
oscillations: (b) dipole mode regime, (c) chaos, and (d) self-trapping of
the excitations. Power spectra are obtained from simulations of a three well
system with $\langle N_j \rangle =10^5$. }
\label{phase_diagram}
\end{figure}
%---------------------------

We next consider the new phenomena that emerge in the strong fragmentation
regime, where the random potential is appreciably larger than the chemical
potential of the atoms so that the tunneling rate for the atoms between the
neighboring wells is much smaller than the local confinement frequency. The
zero temperature condensate wavefunction then can be approximated by the
tight-binding model [\onlinecite{stringari,bishop}]: $\Phi (z)=\sum_{j}\sqrt{%
N_{j}}\phi _{j}(z)\,e^{iS_{j}}$, where $\phi _{j}(z)$ is the localized
single particle wavefunction (here obtained by solving the onsite
Gross-Pitaevskii equation), $N_{j}$ is the number of particles, and $S_{j}$
is the phase of the mini-condensate $j$. Equations of motion of the
mini-condensates (i.e. fragments) are given by 
\begin{eqnarray}
\dot{S}_{j}+U_{j}(N_{j}-N_{j}^{0}) &=&0  \label{eq_S} \\
\dot{N}_{j}+\sum_{\alpha =\pm 1}\tilde{K}_{j,j+\alpha }\sin (S_{j+\alpha
}-S_{j}) &=&0,  \label{eq_N}
\end{eqnarray}%
where $U_{j}$ and $N_{j}^{0}$ are the ``charging energy'' and the
equilibrium number of atoms in the well $j$; $\tilde{K}_{j,j^{\prime
}}\equiv K_{j,j^{\prime }}\sqrt{N_{j}N_{j^{\prime }}}$ is the ``Josephson
energy'' and $K_{j,j^{\prime }}$ is a single particle tunneling rate between
the wells $j$ and $j^{\prime }$. Eq. (\ref{eq_S}) corresponds to the
Josephson relation between the local chemical potential $\delta \mu
_{j}=U_{j}(N_{j}-N_{j}^{0})$ and the phase $S_{j}$. Eq. (\ref{eq_N})
describes the charge conservation.

The key properties of such fragmented BEC can be revealed by exciting BEC in
the ``shaking'' experiments. In these experiments one quickly displaces the
global confining potential along the waveguide and observes the ensuing
motion of the condensate. From the numerical analysis of Eqs. (\ref{eq_S})
and (\ref{eq_N}), we identified three types of response of the system to the
initial displacement, $D$, (see Fig. \ref{phase_diagram}(a)) [%
\onlinecite{endnote1}]: (i) For small $D$ only the dipole mode is excited.
(ii) As $D$ increases, the dipole mode becomes unstable and many other modes
are generated, leading to a chaotic behavior. (iii) When $D$ gets even
larger, the system is ``trapped'' in small oscillations around some
configuration with strong imbalance of the atom density. The number of atoms
in individual wells is not oscillating around the equilibrium values $%
N_{j}^{0}$, but around some non-uniform density distribution. We call this
regime self-trapping of the excitations [\onlinecite{endnote2}].

%-------------------
\begin{figure}[tbp]

\vbox to 5.3cm {\vss\hbox to 5.cm
 {\hss\
   {\includegraphics{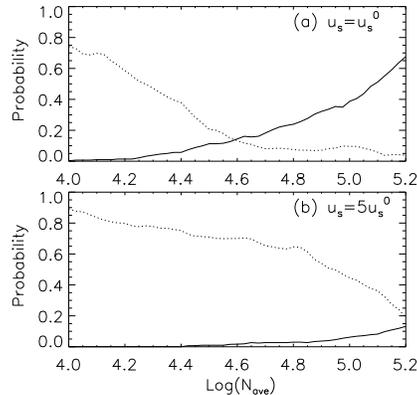}
   }
  \hss}
 }
\caption{The dotted line shows the probability {\cal P}($Q<1$), which is
close to zero if the system is in the superfluid regime and is close to one
if in the insulating Bose glass phase. The solid line shows the probability
that one can observe fluctuations in the number of atoms between the
neighboring wells with $\protect\omega _{J}\geq \protect\omega _{\mathrm{min}%
}$ and $\protect\zeta \geq \protect\zeta _{\mathrm{min}}$ (see text). When
this probability approaches zero, the system will appear self-trapped in the
shaking experiments. $N_{\mathrm{ave}}$ is the average number of atoms in a
single well (mini-condensate). Disorder strength for (a) is the same as used
in Fig. \ref{u_x}(b) for $d=100$ $\protect\mu $m (denoted to be $u_{s}^{0}$), 
while it is five times stronger in (b).}
\label{P_zeta}
\end{figure}
%------------
The power spectra of the oscillations provide a direct way to identify
different types of the dynamical behavior. In the regime (i) condensate
oscillates at the frequency of the dipole mode and its higher harmonics (see
Fig \ref{phase_diagram}(b)). In the regime (ii) we have a broad distribution
of frequencies (see Fig \ref{phase_diagram}(c)). In the limit of
self-trapping, for every point in the system we have several dominant
frequencies and their harmonics (see Fig \ref{phase_diagram}(d)). The
transition of the system from a dipole mode into a chaotic regime originates
from the non-linearity of Eqs. (\ref{eq_S})-(\ref{eq_N}) and is similar to
the modulation instability in optical lattices [\onlinecite{bishop}]. It is
commonly believed that for strong disorder the superfluidity are terminated
by strong quantum fluctuations and that the system should go into the
insulating Bose glass phase (BG) [\onlinecite{giamarchi_and_fisher}].
Therefore, the coherent dipole fluctuations should disappear at the SF/BG
transition point, and the self-trapping response should become dominant in
the BG phase due to the localized nature of the insulating state. The
qualitative phase diagram is shown in Fig. \ref{phase_diagram}(a).

More quantitatively, transition into an insulating state can be estimated by
the ratio of the ``Josephson energy'' between the neighboring wells to the
average charging energy, $Q_{j}=\frac{8\tilde{K}_{j,j+1}}{(U_{j}+U_{j+1})/2}$%
. The SF to the Mott insulator transition in a periodic potential takes
place when $Q=1$in the meanfield theory. For a correlated random potential
we expect that the SF to the BG phase transition takes place when the
probability to have $Q<1$ in a junction, $\mathcal{P}(Q<1)$, is of the order
of one. In Fig. \ref{P_zeta} we show $\mathcal{P}(Q<1)$ as a function of the
atom density for two different strengths of the disorder potential (dotted
lines). Taking the parameters corresponding to the experiments of Ref. [%
\onlinecite{aaron1}] with $d=100$ $\mu $m, we estimate that a system of
atoms with $N_{\mathrm{ave}}<10^{4}$ per fragment should be in the BG phase.

Naturally, the shaking experiments can be used to identify the SF/BG
transition. In practice, a finite lifetime of a condensate limits
experimentally measurable Josephson frequencies $\omega _{J}$ to be $\omega
_{J}\geq \omega _{min}$. When a typical $\omega _{J}$ becomes smaller than $%
\omega _{\mathrm{min}}$, oscillations may no longer be detected. Another
limiting factor for the experiments comes from the finite density contrast, $%
\zeta =\frac{\Delta N}{N_{0}}$, required for resolving oscillations. When
typical $\zeta $ become smaller than $\zeta _{\mathrm{min}}$ (for a fixed $D$%
), the system will always appear trapped. The solid curve in Fig. \ref%
{P_zeta} shows the probability of observing a dipole mode or chaotic
dynamics for a given disorder strength with the minimum displacement for $%
\omega _{\mathrm{min}}=2\pi \cdot 1$Hz and $\zeta _{\mathrm{min}}=0.1$. 
The observation of the self-trapping transition as a
function of the disorder strength at a small $D$ should provide a reasonable
estimate of the SF-BG transition point [\onlinecite{endnote1}], as
illustrated by the dashed line in Fig. \ref{phase_diagram}(a). We expect
such dynamical properties and a sharp crossover to the insulating regime are
still valid, at least qualitatively, in the temperature regime of current
experiments.

In the shaking experiments discussed so far the random potential was kept
fixed and the confining potential was displaced. It would also be
interesting to keep the confining potential fixed and oscillate the random
magnetic potential at some small frequency. This would be a direct analogue
of the Andronikashvilli experiments for superfluid $^{4}\mathrm{He}$ in a
disordered medium [\onlinecite{reppy_review}], which provide a direct
measure of the superfluid condensate fraction.

In summary, we performed microscopic calculations of a random magnetic
potential for the atomic BEC in a microtrap. We showed that a small
meandering of the wire is sufficient to explain the experimentally observed
fragmentation of the condensates [\onlinecite{note0}]. The response to the
trap shaking can be used to study the superfluid properties of the strongly
fragmented condensates and to identify the quantum phase transition from
superfluid to the insulating Bose glass phase. More generally, our work
demonstrates that atoms in the magnetic microtrap is a very promising system
for controlled study of one-dimensional disorder problems.

%%%%%%%%%%%%%% acknowledgement %%%%%%%%%%%%%
The authors acknowledge useful discussion with E. Altman, E. Heller, D.
Nelson, M. Prentiss, and D.E. Pritchard. We especially thank A.E. Leanhardt
and S. Kraft for explaining details of the experiments. This work was
supported by the NSF (grants DMR-01328074, PHY-0134776), the Sloan and the
Packard Foundations, and by Harvard-MIT CUA.

%%%%%%%%%%%%%%%%%%%%%%%%%%%%%%%%%%%%%%%%%%%%%
%%%%%%%%%%%%%%%%%%%%%%%%%%%%%%%%%%%%%%%%%%%%%

%%%%%%%%%%%%%%%%%%%%%%%%%%%%%%%%%%%%%%%%%%%%%%%%

%%%%%%%%%%%%%%%%%%%%%%%%%%%%%%%%%%%%%%%%

%%%%%%%%%%%%%%%%%%%%%%%%%%%%%%%


\begin{thebibliography}{99}
\bibitem{henkel} The thermal fluctuations from the substrate surface are
negligible in the present atom-wire distance.
\bibitem{note0} We note that such disorder effects can be neglected in the
macroscopic magnetic traps, where the atoms are far away from the wire ($%
d\sim 1$ cm) and hence the disorder vanishes according to Eq. (\ref{u_s}).
\bibitem{endnote1} Note that in the thermodynamic limit, various regimes may
be distinguished only for a bounded disorder, while in a finite system there
is no true phase transition --- however, one may still observe the sharp
crossovers.
\bibitem{endnote2} Note that the ends of the condensate may be trapped even
when the bulk of the system is not.
\end{thebibliography}
\end{document}